\documentclass[aps,prl,letter,twocolumn,superscriptaddress,floatfix,nofootinbib,showpacs,amsmath,amssymb,altaffilletter]{revtex4}
\usepackage{graphicx}
\usepackage{dcolumn}
\usepackage{bm}
\usepackage{color}
\usepackage{verbatim}
\usepackage{paralist}
\usepackage{amssymb}
\usepackage[english]{babel}
\usepackage{color}
\newcommand{\ops}{$\emph{o-Ps}$}
\newcommand{\ps}{$\emph{Ps}$}
\newcommand{\pps}{$\emph{p-Ps}$}

\begin{document}

\title{Positronium signature in organic liquid scintillators for neutrino experiments}

\newcommand{\Polimi}{Dipartimento di Fisica, Politecnico di Milano, p.zza Leonardo da Vinci 32, 20133 Milano, Italy}
\newcommand{\Milano}{Dipartimento di Fisica, Universit\`a and INFN Milano, via Celoria 16, 20133 Milano, Italy}
\newcommand{\APC}{Astroparticule et Cosmologie APC, 10 rue Alice Domon et Leonie Duquet, 75205 cedex 13, Paris, France}

\author{D.~Franco}\thanks{davide.franco@apc.univ-paris7.fr }\affiliation{\APC}
\author{G.~Consolati}\affiliation{\Polimi}
\author{D.~Trezzi}\affiliation{\Milano}
%%%%%%%%%%%%%%%%%%%%%%%%%%%%%%%%%%%%%%%
\date{\today}

\begin{abstract}

Electron anti-neutrinos are commonly detected in liquid scintillator experiments via inverse beta decay, by looking at the coincidence between the reaction products, neutron and positron. Prior to positron annihilation, an electronÐ-positron pair may form an orthopositronium (\ops) state, with a mean life of a few ns. Even if the \ops\ decay is speeded up by $\emph{spin flip}$  or $\emph{pick off}$ effects, it may introduce distortions in the photon emission time distribution, crucial for position reconstruction and pulse shape discrimination algorithms in anti-neutrino experiments. Reversing the problem, the  \ops\  induced time distortion represents a new signature for tagging anti-neutrinos in liquid scintillator.

In this paper, we report the results of measurements of the \ops\ formation probability and lifetime, for the most used solvents for organic liquid scintillators in neutrino physics (pseudocumene , linear alkyl benzene,  phenylxylylethane, and dodecane). We characterize also a mixture of pseudocumene +1.5 g/l of 2,5-diphenyloxazole, a fluor acting as wavelength shifter.  

In the second part of the paper,  we demonstrate that the  \ops\ induced distortion of the scintillation photon emission time distributions  represent an optimal signature for tagging positrons on an event by event basis, potentially enhancing the anti-neutrino detection.

\end{abstract}


%\end{keyword}

\maketitle

\section{Introduction}

Electron anti-neutrinos are produced in $\beta$ decays  of naturally occurring radioactive isotopes in the Earth, representing a unique direct probe of our planet's interior. Recent results by  Borexino \cite{BX10} and KamLAND \cite{KAM05} have been combined to  provide the first observation  of geo-neutrinos at 5 $\sigma$ \cite{Fog10}. Also nuclear reactors provide intense sources of antineutrinos, which come from the decay of neutron-rich fragments produced by heavy element fissions. In the past, reactor anti-neutrinos  allowed to constrain the neutrino oscillation parameters in the Solar sector \cite{KAM03}. Presently, several experiments (Double Chooz \cite{DC06}, Daya Bay \cite{DB07}, RENO \cite{REN07}) exploit reactor anti-neutrinos to investigate the last unknown neutrino oscillation parameter, $\Theta_{13}$. Moreover, reactor anti-neutrinos are directly linked to the fissile isotope production and consumption processes and can be studied for monitoring the nuclear plants \cite{BER02}. 

One of the main advantages in detecting electron anti-neutrinos is the clean signature provided by the inverse beta decay reaction:
\begin{equation}
\bar{\nu}_e + p \rightarrow n + e^+
\label{eq:inverse}
\end{equation}

by looking at the neutron-positron coincidence. 

In organic liquid scintillators, neutrons are  captured mainly on protons, and identified looking at the characteristic 2.223 MeV $\gamma$-ray emitted in the reaction:
\begin{equation}
n + p \rightarrow d + \gamma
\end{equation}

To enhance the neutron detection, scintillators can be  loaded with  high neutron capture cross section materials, like Gadolinium. The neutron mean capture time varies from few tenths up to hundreds of microseconds. 

Positrons interacting in liquid scintillator may either directly annihilate with electrons or form positronium (\ps) \cite{Deu51}, a bound state of a positron and an electron with two sub-states: 25\% of positronium populates the spin singlet state (S=0), called para-positronium (\pps), and 75\% the spin triplet state (S=1), called ortho-positronium (\ops). \pps\ annihilates emitting two $\gamma$-rays of 511 keV with a mean life of  125 ps in vacuum.  \ops\ emits three $\gamma$-rays, with a total energy equal to twice the electron mass, and a mean life in vacuum equal to $\sim$140 ns. 

In matter, however, interactions of \ops\ with the surrounding medium  strongly reduce its lifetime: processes like chemical reactions, \emph{spin-flip} (ortho-para conversion at paramagnetic centers), or pick-off annihilation on collision with an anti-parallel spin electron, lead to the two body decay with lifetimes of a few nanoseconds. The surviving three body decay channel is typically  reduced to a negligible fraction. 

If the delay introduced by the positron annihilation lifetime is of the order of  a few  nanoseconds, calorimetric scintillation detectors, like Borexino \cite{BOR08} and KamLAND \cite{KAM08},  are unable to disentangle the energy deposited by  positron interactions from that released by annihilation $\gamma$-rays.  In these cases, a delayed  $\gamma$-ray emission induces a distortion in the time distribution of detected photoelectrons (\emph{pulse shape}), with respect to a direct annihilation event. Such a distortion can affect algorithms based on the pulse shape, like position reconstruction and particle discrimination. 

\begin{table}[h]
\begin{center}
\begin{tabular}{llll} 
Experiment & Scintillator & Fluor & Dope  \\
\hline 
\hline 
KamLAND \cite{KAM03}  & 20\% PC & 1.5 g/l PPO \\
                           &                     80\% OIL                  &&\\

Borexino \cite{BOR08}     & PC & 1.5 g/l PPO \\

LVD \cite{LVD89} & Paraffin& 1.0 g/l PPO \\

SNO+ \cite{SNO08} & LAB & PPO & 0.1\% Nd \\

Double Chooz \cite{DC06} & 20\% PXE &3-6 g/l PPO  & 0.1\% Gd\\
                           &                     80\% OIL                  &20 mg/l Bis-MSB&\\

Daya Bay \cite{DB07} & LAB & 3 g/l PPO & 0.1\% Gd \\
                   &         &15 mg/l Bis-MSB&\\

RENO \cite{REN07}&  LAB & 1-5 g/l PPO & 0.1\% Gd\\
                   &         &1-2 mg/l Bis-MSB&\\
\hline
\end{tabular}
\end{center}
\caption{Scintillator composition used in present and future underground neutrino experiments.}
\label{tab:exp}
\end{table}%

Distortions in positron pulse shapes  affect not only anti-neutrino detection via inverse beta decay, but also  the reconstruction of $\beta^+$ decay events, like those from  cosmogenic $^{10}$C and $^{11}$C radioisotopes, which represent  a crucial background for the Solar neutrino signal in Borexino and KamLAND \cite{BX06, Gal05}. 

In this paper, we characterize the \ops\ formation probability and lifetime for the most popular choices of  scintillator solvents used by present experiments (table    \ref{tab:exp}): 1,2,3 trimetilbenzene or pseudocumene  (PC, C$_{9}$H$_{12}$), linear alkyl benzene (LAB, C$_{18}$H$_{30}$),  phenylxylylethane (PXE, C$_{16}$H$_{18}$) and dodecane (OIL, C$_{12}$H$_{26}$).  

Wavelength shifters are present in scintillators with very low concentrations. However, adding a fluor to scintillators  can change the \ops\ properties. For this reason,  we studied a typical scintillator mixture, PC + 1.5 g/l of PPO (2,5-diphenyloxazole), to observe the effect of the fluor on the \ops\ properties.

In the second part of the paper we discuss the results of Monte Carlo simulations in which we quantify the distortion on the pulse shapes introduced by the \ops\ and the effects on the positron event reconstruction in large volume liquid scintillator detectors.

Finally, we demonstrate that the \ops\ induced distortion potentially represents a powerful signature for anti-neutrinos.

\section{Experimental apparatus}
The positron annihilation lifetime spectra have been measured with a fast coincidence system, formed by two plastic scintillators, with a time resolution of 0.28 ns at FWHM.  $^{22}$Na is a $\beta^+$ radioisotope, with  Q-value = 2.842 MeV, emitting positrons in association with   1.274 MeV $\gamma$-rays  (99.94\% branching ratio).  The $^{22}$Na  source  (0.8 MBq activity) was prepared by drying a droplet of $^{22}$Na from a carrier free neutral solution between two layers, each composed of two Kapton foils (7.5 $\mu$m thick, 1 cm radius each single layer).  Kapton is a polymide compatible with scintillator materials and where positrons do not form \ops\ states \cite{Mcg06, Plo88}. 

The Kapton--source sandwich is inserted in a glass vial, containing the scintillator sample. The vial is positioned between two plastic scintillator detectors (Pilot U) each of them coupled with a photomultiplier tube. The vial is in direct contact with the faces of the two detectors, in order to maximize the solid angle.  

The system measures the delay occurring between the $^{22}$Na decay and the positron annihilation: when a plastic scintillator is hit by a 1.274 MeV $\gamma$, from $^{22}$Na decay,  the second detector opens a gate of 43 ns in order to detect one of the annihilation $\gamma$'s. The time resolution of each detector is $\sim$ 130 ps.

Discrimination between start and stop events occurs on the basis of the different energies of the detected $\gamma$-ray.  The first detector has a lower threshold at 900 keV to detect the prompt signal from the 1.274 MeV $\gamma$, while the annihilation gamma deposited energy  is recorded if it falls in the [350--500] keV energy range. The second energy cut has been optimized to avoid the back scattering of the 1.274 MeV $\gamma$'s. 

\begin{figure}[]
\begin{center}
\includegraphics[scale=0.45]{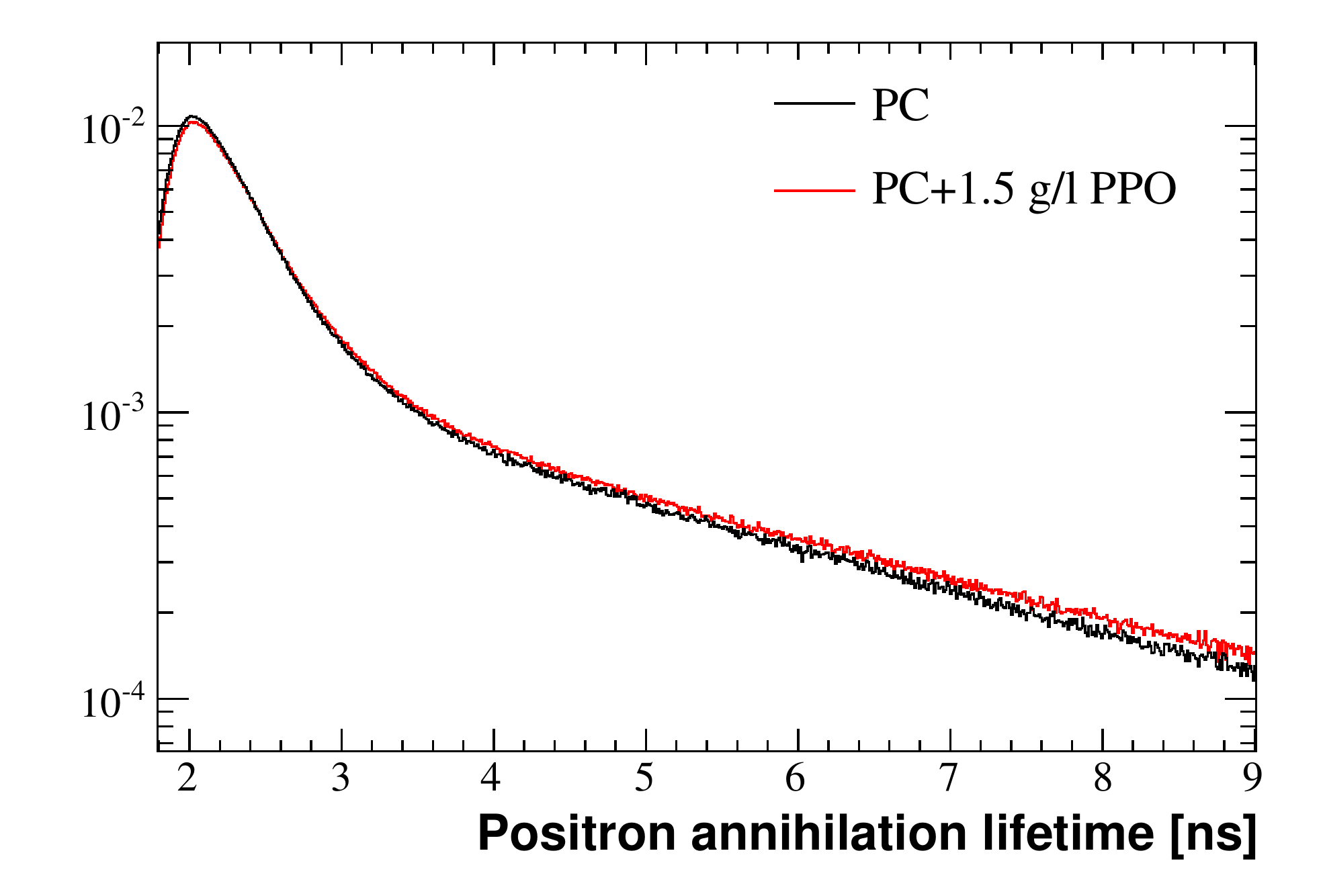}
\caption{Color online. Positron annihilation lifetime spectra for the PC and PC+1.5 g/l PPO samples.}
\label{fig:spectra}
\end{center}
\end{figure}

A constant fraction discriminator generates a fast timing signal whenever a $\gamma$-ray with the correct energy is detected. A time-to-amplitude converter, enabled by the start signal from the discriminator, produces a voltage linearly increasing with time, which stops at the arrival of the stop signal from the other discriminator: the signal at the output of the time-to-amplitude converter is proportional to the time interval elapsed between generation and annihilation of the positron. This signal is finally digitized by an analog-to-digital converter with 4096 channels.

The time interval corresponding to each channel is 10.6 ps, and it has been calibrated with a  $^{60}$Co source. The resolution of the apparatus (0.28 ns at FWHM) is measured by looking at the fast (182 ps) coincidence between the two $\gamma$'s emitted by the  $^{207}$Bi decay, through the $5/2^-$ $^{207}$Pb excited state. The  detector resolution is  modeled with the sum of two gaussians with different standard deviations.

\section{Data analysis and results}

The 5 samples   (PC, PXE, LAB, OIL, PC+PPO) of scintillators have been degassed with nitrogen before each measurement. For each sample, we repeated the measurement, at the ambient temperature, from 3 to 5  times, to take into account  possible systematics due to changes of the initial conditions (e. g. temperature and time of nitrogen stripping). 

The  collected total statistics for each sample is  $\sim5\times10^6$  events. Measured positron annihilation lifetime distributions of PC and PC+PPO samples are shown in figure \ref{fig:spectra}.

\begin{figure}[]
\begin{center}
\includegraphics[scale=0.45]{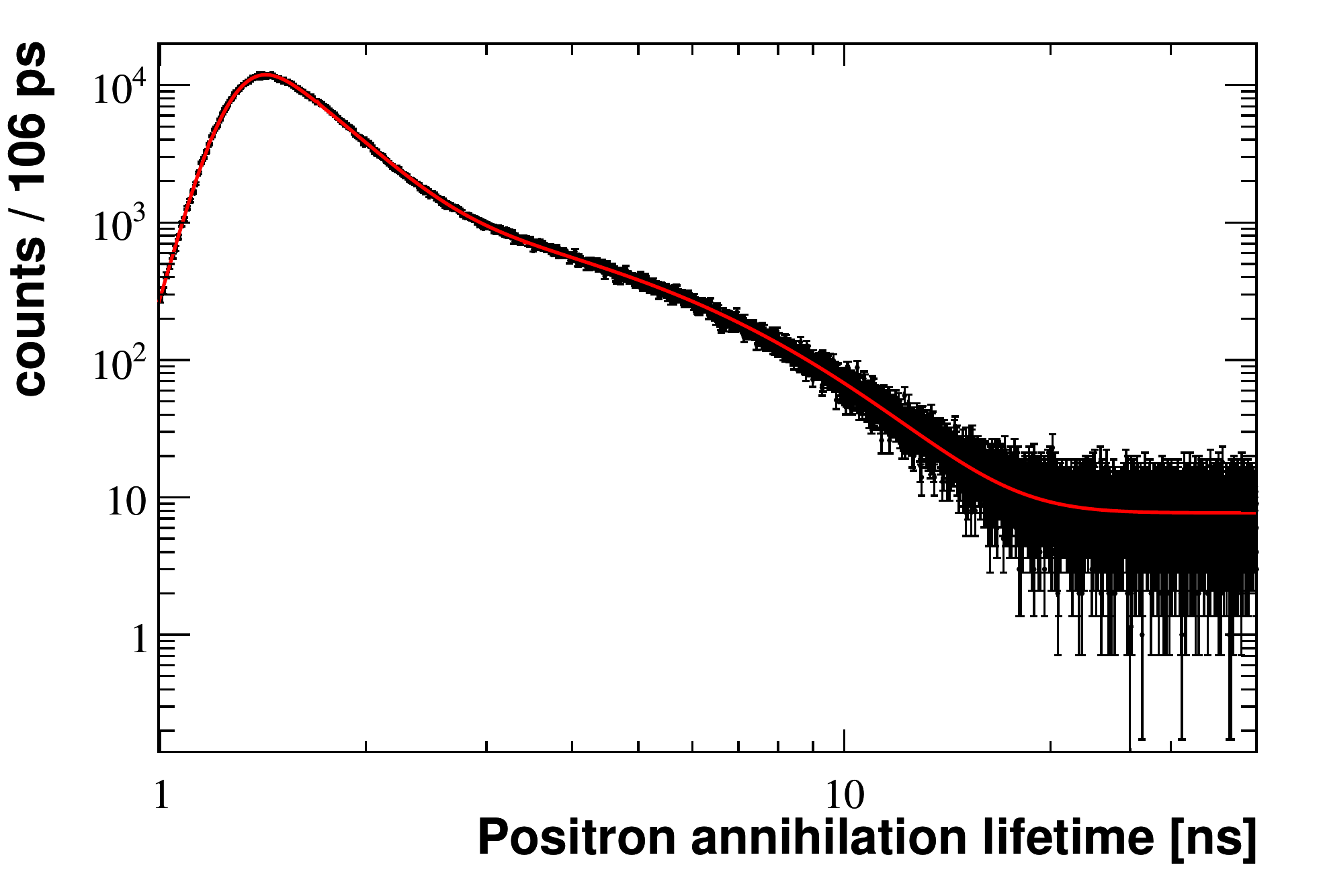}
\caption{Color online. Fit (red line) of the positron annihilation life time spectrum (black dots) for the PXE sample.}
\label{fig:fit}
\end{center}
\end{figure}

Each distribution is fitted with a three component model:

\begin{equation}
F(t) =  \chi(t > t0)\cdot \left(\sum_{k=1,2}  \frac{A_k}{\tau_k} \cdot e^{-t/\tau_k} + C\right)
\label{eq:ft}
\end{equation}

where A$_1$ and $\tau_1$   correspond to the effective amplitude and mean life  of the annihilation and \pps\ components, which can not be disentangled; A$_2$ and $\tau_2$  are the  \ops\ amplitude and mean life, respectively; $C$ is  the flat distribution due to  accidental background; $t_0$ is the detector time offset and $\chi$ is a step function. The fit function $F(t)$ is convoluted with the detector resolution, modeled with the sum of two gaussians:

\begin{equation}
G(t) =  \sum_{i=1,2} \frac{g_i}{\sqrt{2\pi \sigma_i^2}}\cdot e^{-\frac{t^2}{2\cdot\sigma_i^2}}
\end{equation}
 
centered in the same value but with different resolutions ($\sigma_1$ and $\sigma_2$), and where $g_i$ ($i=1,2$) are the relative weights of the two components ($g_1 + g_2 =1$).   The detector resolution is dominated by the  gaussian with  $\sigma_1$ $\sim$ 110 ps ($g_1$ $\sim$ 0.8),  while $\sigma_2$ $\sim$ 160 ps. The data modeling package used in this analysis is the RooFit toolkit \cite{Ver03}, embedded in the ROOT package \cite{Bru97}, based on MINUIT. All the parameters in the model are free in the fit. An example of fit for the PXE sample is shown in figure \ref{fig:fit}.

All the normalized $\chi^2$, obtained from the fits, are in the [0.85--0.98] range. Statistical errors for both mean lives and amplitudes are at the order of per mil level. $\tau_1$ is almost constant in all the measurements: the mean $\tau_1$ is centered in 365 ps, with a root mean square of 8 ps.

\begin{figure}[]
\begin{center}
\includegraphics[scale=0.45]{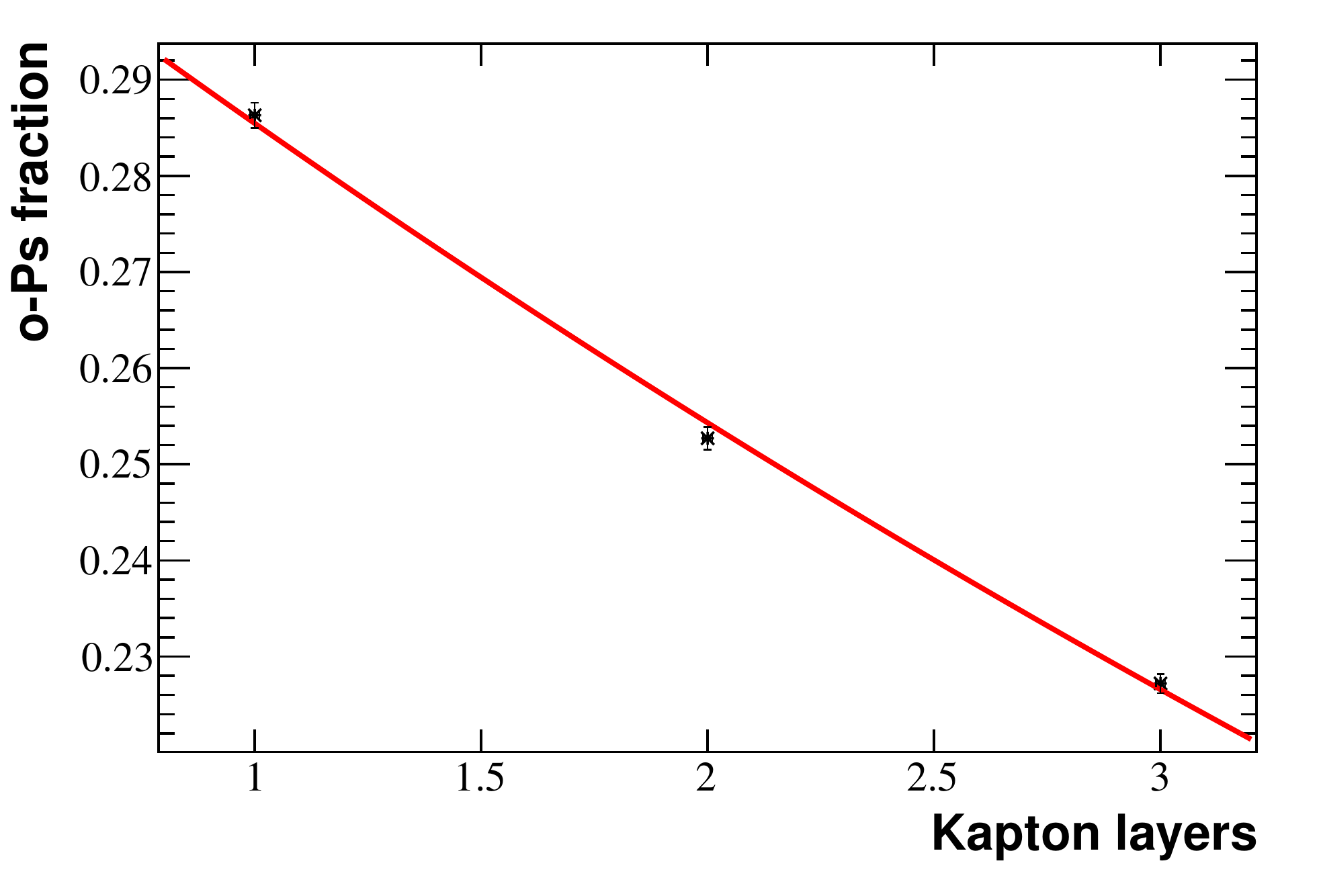}
\caption{Color online. The data points represent the \ops\ fraction, measured with  1 to 3  Kapton layers, inserted in a Plexiglas medium. Each layer is 7.5 $\mu$m thick and 1 cm radius. The line is  the result of an exponential fit.}
\label{fig:kapton}
\end{center}
\end{figure}

To estimate the fraction of positrons annihilating in Kapton, we  covered the $^{22}$Na source with 1 to 3 Kapton layers, identical to those used for charactering scintillators.  The Kapton--source sandwiches  were inserted in a Plexiglas medium, characterized  by an  \ops\  mean life of $\sim$ 2 ns. The  \ops\ fraction, as function of the  Kapton layers,  is fitted with an exponential, as shown in figure   \ref{fig:kapton}.  We estimate that  20.6 $\pm$ 0.2\% of positrons annihilate in the double Kapton layer (15 $\mu$m thick).  

In order to derive the \ops\ formation probability ($f_2$) in the analyzed  liquid scintillator samples, we subtract the annihilation component in Kapton ($A_K$):

\begin{equation}
f_2 = \frac{A_2}{A_1 + A_2 - A_K}.
\end{equation}

\begin{figure}[]
\begin{center}
\includegraphics[scale=0.45]{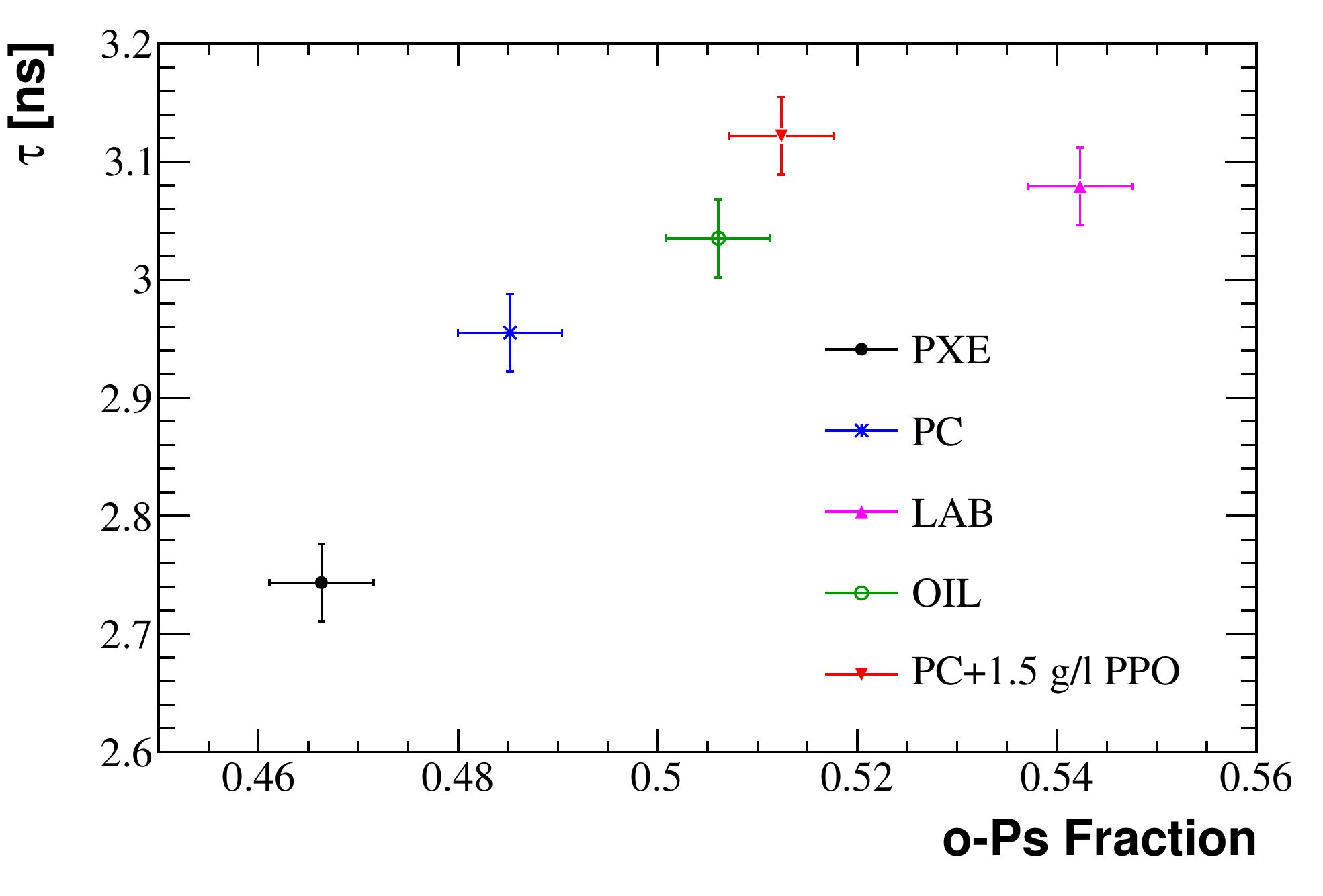}
\caption{Color online. Results of the \ops\ probability formation and mean life for the analyzed samples of scintillators.}
\label{fig:res}
\end{center}
\end{figure}

To estimate the systematic uncertainties, we evaluated first the  \ops\  mean life weighted averages of all the measurements for each sample.  We fitted then the distribution of all the deviations of each measurement from the correspondent weighted average with a gaussian, in a likelihood approach. The same procedure has been applied also for the  \ops\  formation probability.  The resulting systematic errors for the \ops\ fraction and mean life are 0.5\% and 0.03 ns, respectively, and dominate over the statistical errors. Final results are shown in table \ref{tab:res} and in figure \ref{fig:res}.

All the samples analyzed in this paper are characterized by an \ops\ mean life about 3 ns and a formation probability about 50\%. The  \ops\  mean life  and fraction for PXE and LAB  represent the extremes in the value ranges: PXE (LAB) is the solvent where \ops\ has the lowest (largest) probability to be formed and the shortest (longest) life time.  

Kino et al.  \cite{Kin00}  characterized the \ops\ properties  on a PC sample saturated with N$_2$  with a similar experimental setup. The resulting mean life (3.03 $\pm$ 0.02 ns) is compatible within 2 $\sigma$ with the one reported in this paper, while we observe a formation probability  5.2\% larger.   This small discrepancy can be attributed  to a not identical concentration of oxygen \cite{Kin00}  in the scintillator sample, or to different  condition parameters, like temperature (the \ops\ formation probability as function of the temperature has been discussed by B. Zgardzinska et al.  \cite{Zga06}), during the measurements.

\begin{table}[b]
\begin{center}
\begin{tabular}{lcc} 
Material & $f_2$ & $\tau_2$ [ns] \\
\hline 
\hline 
PXE & 0.466                     $\pm$ 0.005 & 2.74 $\pm$ 0.03 \\
LAB & 0.542                     $\pm$ 0.005 & 3.08 $\pm$ 0.03\\
PC  & 0.485                      $\pm$ 0.005 & 2.96 $\pm$ 0.03\\
OIL & 0.506                       $\pm$ 0.005 & 3.04 $\pm$ 0.03\\
PC+1.5 g/l PPO & 0.512 $\pm$ 0.005 & 3.12 $\pm$ 0.03\\
\hline
\end{tabular}
\end{center}
\caption{Final results for the \ops\ probability formation and mean life for the analyzed samples of scintillators.}
\label{tab:res}
\end{table}

\section{\ops\ induced pulse shape distortion}

Positrons and the following annihilation gammas, even after \ops\ formation, can not be disentangled in large volume liquid scintillator detectors, like Double Chooz, Borexino and KamLAND, by looking at the photon arrival times in the photomultiplier tubes. Such experiments are, in fact, characterized by time constants, like the   fluorescence decay-time (table \ref{tab:scint}) and the photomultiplier tube time jitters, typically longer than the delay between positron and annihilation gamma emission.

\begin{figure}[t]
\begin{center}
\includegraphics[scale=0.45]{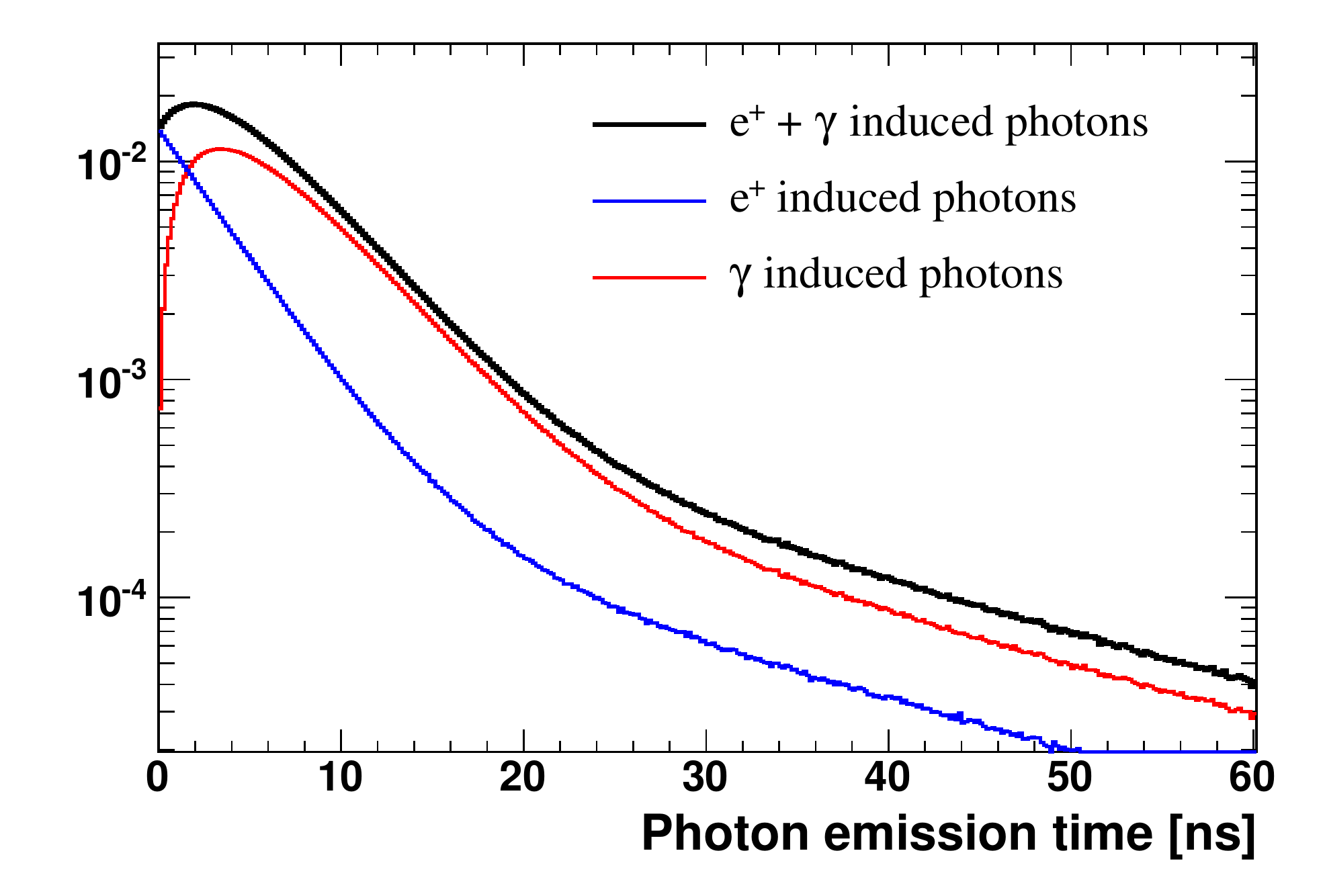}
\caption{Color online. Simulation of the photon emission time (black line) induced by 0.5 MeV positrons, all forming \ops, in the PC+1.5 g/l PPO scintillator. The  scintillation  components, due to positron (blue) and \ops\ decay $\gamma$'s (red), are shown separately.}
\label{fig:pulse}
\end{center}
\end{figure}

However, the   \ops\ formation can induce a significant distortion in the   pulse shape of positrons. In the positron detection, scintillator molecules are first excited by positron interactions, and then by annihilation $\gamma$--rays.  If annihilation passes through the intermediate \ops\ state, the annihilation component is delayed, and the overall photon emission time distribution (PETD) results as the sum of the two components, as shown in figure \ref{fig:pulse} for 0.5 MeV positrons in PC + 1.5 g/l PPO,  annihilating after  \ops\ formation. The correspondent  PETD deformation, with respect to the direct positron annihilation case, is  dominant in the first 30 ns.

\begin{table}[b]
\begin{center}
\begin{tabular}{lllllll}
Scintillator  & $\tau_1$ & $\tau_2$ & $\tau_3$& N$_1$& N$_2$& N$_3$ \\
                     & [ns]          & [ns] & [ns]   & \% &  \% &  \% \\
\hline
\hline
PC + 1.5 g/l PPO & 3.57 & 17.61 & 59.9 & 89.5 &  6.3 &  4.2 \\
PXE + 1.0 g/l PPO & 3.16 & 7.7 & 34 & 84.0 &  12.0 & 2.9 \\
LAB + 1.0 g/l PPO & 7.46& 22.3& 115 &75.9 & 21.0 & 3.1 \\
\hline
\end{tabular}
\end{center}
\caption{Scintillator decay time  constants ($\tau_i$) and amplitudes (N$_i$) for $\beta$  particles for PC+1.5g/l PPO \cite{Bac08}, PXE + 1.0 g/l PPO \cite{Mar09}, LAB + 1.0 g/l PPO \cite{Mar09}}
\label{tab:scint}
\end{table}%

\begin{figure}[]
\begin{center}
\includegraphics[scale=0.45]{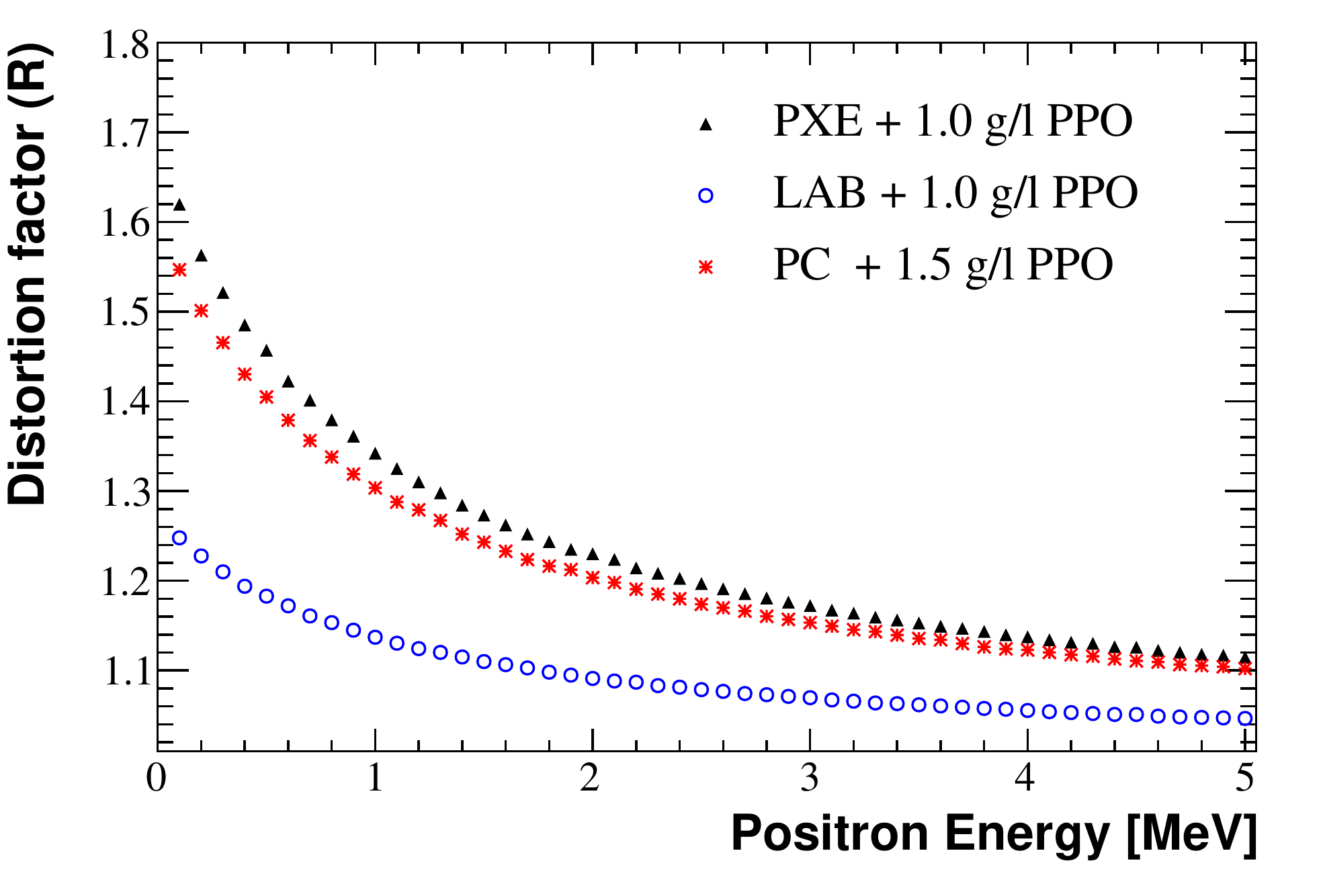}
\caption{Color online. Ratio between the mean values of the \ops\ and direct annihilation photon emission time distributions as function of the positron energy for PXE + 1.0 g/l PPO, LAB + 1.0 g/l PPO, and PC + 1.5 g/l PPO, assuming the scintillation decay constants from table \ref{tab:scint} }
\label{fig:mean}
\end{center}
\end{figure}

The distortion is energy dependent, since the amplitude of the first component is proportional to the positron energy, while the gamma component has fixed energy. To study such dependency, we simulated positrons with energies from 0.1 to 5.0 MeV.  The ratio ($R$) of the PETD  mean values between the \ops\ and the direct annihilation cases, is  plotted in figure \ref{fig:mean}. It is notable that the deformation of the time distribution is larger for lower energy positrons, where $R$ can reach up to $\sim$1.6. At higher energies, the distortion reaches a plateau, with \ops\ PETD mean values  shifted up to 10\%, depending on the scintillator composition.

In neutrino experiments, the \ops\ induced PETD deformation can have an impact  on several algorithms for the  event reconstruction and discrimination. For instance,  the position reconstruction algorithms (e.g. O. Smirnov \cite{Smi03})  strongly depends on the first (t$<$30 ns) detected photoelectrons. A bias in the position reconstruction can induce a further bias in the event energy reconstruction, since the light collection on the photomultiplier tubes depends on the reconstructed position. Also algorithms  for particle discrimination  in liquid scintillators can be strongly affected, since they  rely on the dependence of fast and slow portions of the scintillation pulse  on the energy loss of the interacting particle  \cite{Gat62, Bac08}.

\section{Positronium signature}

Beyond the necessity to minimize systematics in the event reconstruction, the \ops\ induced pulse shape distortion potentially provides a signature  for tagging positrons, and hence for enhancing the neutrino detection. Borexino and KamLAND could directly benefit by a positron/electron discrimination by enhancing the tagging and rejection of cosmogenic $^{11}$C $\beta^+$decay, the main background component in the $pep$ and CNO solar neutrino energy window \cite{BX06, Gal05}. Also $\theta_{13}$ and geo-neutrino experiments  could be advantaged by discriminating cosmogenic $^9$Li and $^8$He $\beta$-neutron decays which mimic the anti-neutrino signal. 

In order to quantify the \ops\ and the direct annihilation positron event discrimination, we simulate an ideal spherical detector (4 m radius), with 2000 photomultiplier tubes. The simulation is Geant4 based. The cathode is a segment of a  20 cm radius sphere. The scintillator  is PC+1.5 g/l PPO, and its  properties  are taken from F. Elisei et al.  \cite{Eli97}. The scintillation photon yield is 10,000 photons/MeV. The simulation includes optical effects, like Rayleigh scattering, absorption and re-emission, and reflections on cathodes and on the spherical steel structure. 

\begin{figure}[]
\begin{center}
\includegraphics[scale=0.45, angle=90]{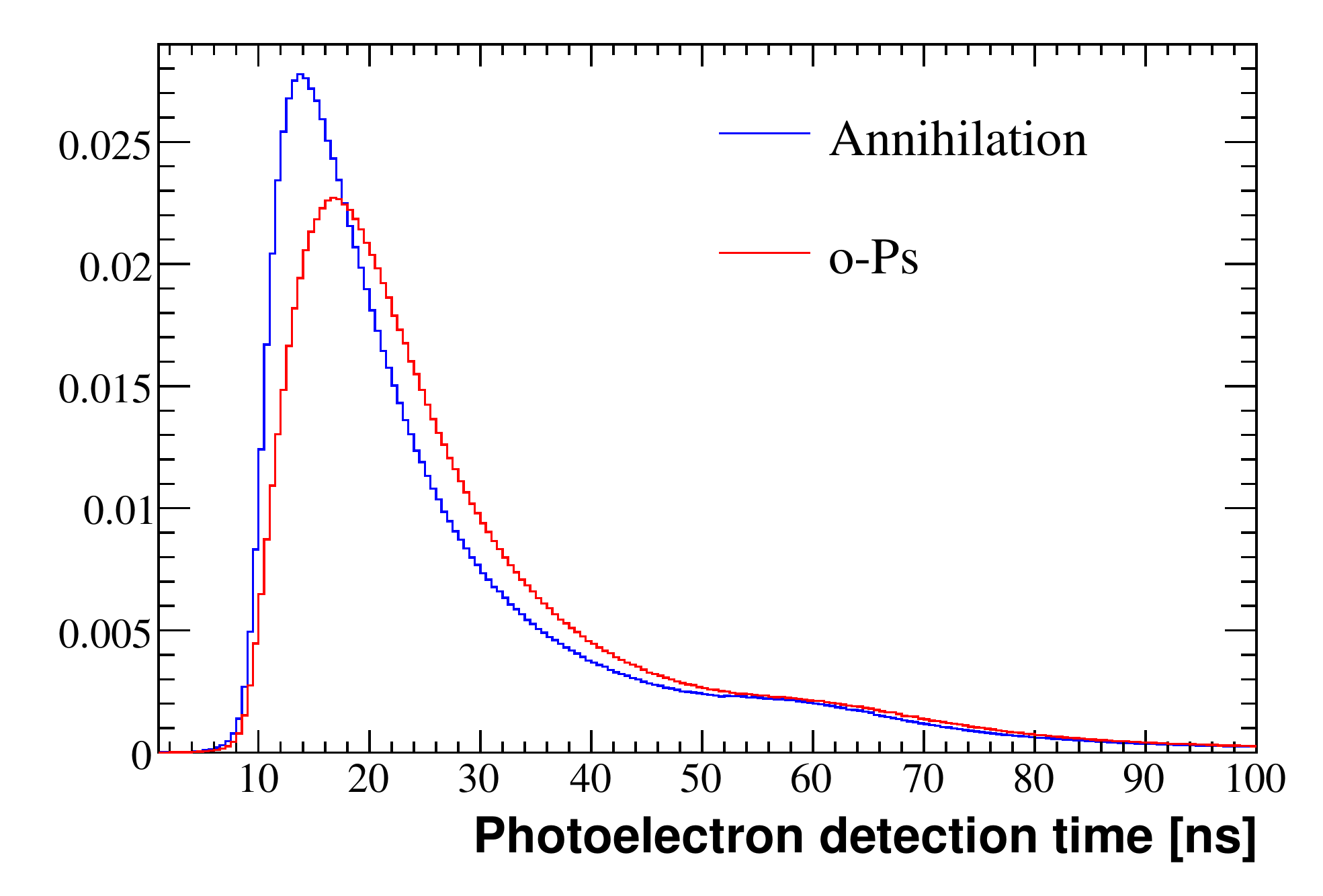}
\caption{Color online. Photoelectron time distributions from an ideal 4 m radius detector, filled with PC+1.5 g/l PPO, for 0.5 MeV positrons directly annihilating (blue) and forming \ops\ (red). The structure at $\sim$60 ns is due to optical reflections on detector materials. Both the distributions are normalized to 1 event.}
\label{fig:tof}
\end{center}
\end{figure}

We simulated samples of 10,000 positron events, in the detector center, directly annihilating or annihilating following the \ops\ formation. Positrons, directly annihilating, have the identical pulse shapes of equivalent electrons with same energies plus 1.022 MeV, from  annihilation gammas. Positrons, in fact, thermalize in ~300 ps, much faster than the characteristic scintillation decay times, and gammas release energy mainly through Compton electron cascades.  In the simulation, positron energy varies from 0.1 to 5 MeV. To take into account the photomultiplier tube jitter, we smeared the photoelectron time distribution  by 1.4 ns and we assume an ideal fast electronic chain based on  1 GHz Flash--ADC. The photoelectron time distributions (figure \ref{fig:tof}) are relative to the first detected photoelectron. 

\begin{figure}[]
\begin{center}
\includegraphics[scale=0.45, angle=90]{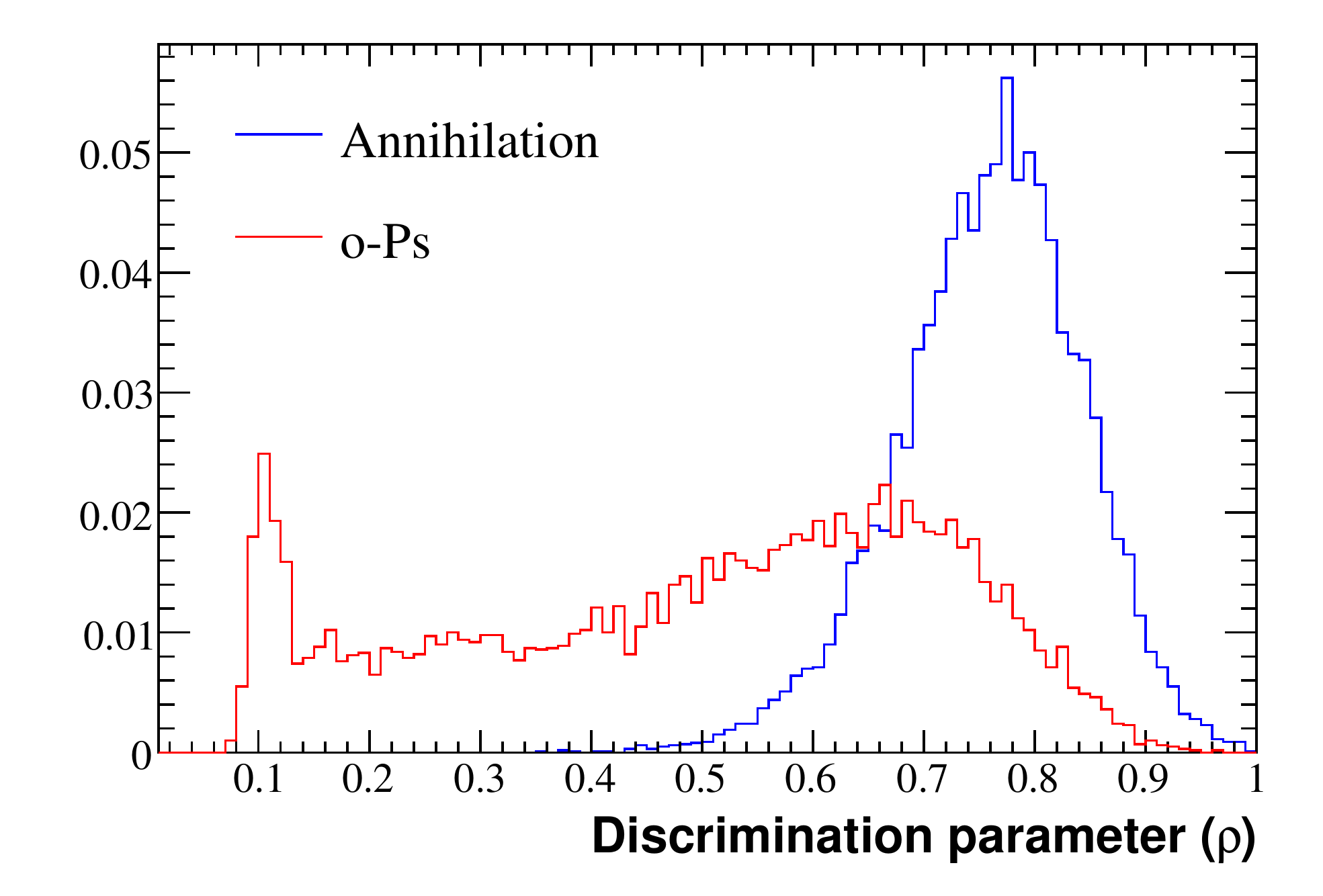}
\caption{Color online. $\rho$ distributions for 0.5 MeV positrons directly annihilating (blue) and forming \ops\ (red). Both the distributions are normalized to 1 event.}
\label{fig:rho}
\end{center}
\end{figure}

\begin{figure}[]
\begin{center}
\includegraphics[scale=0.45]{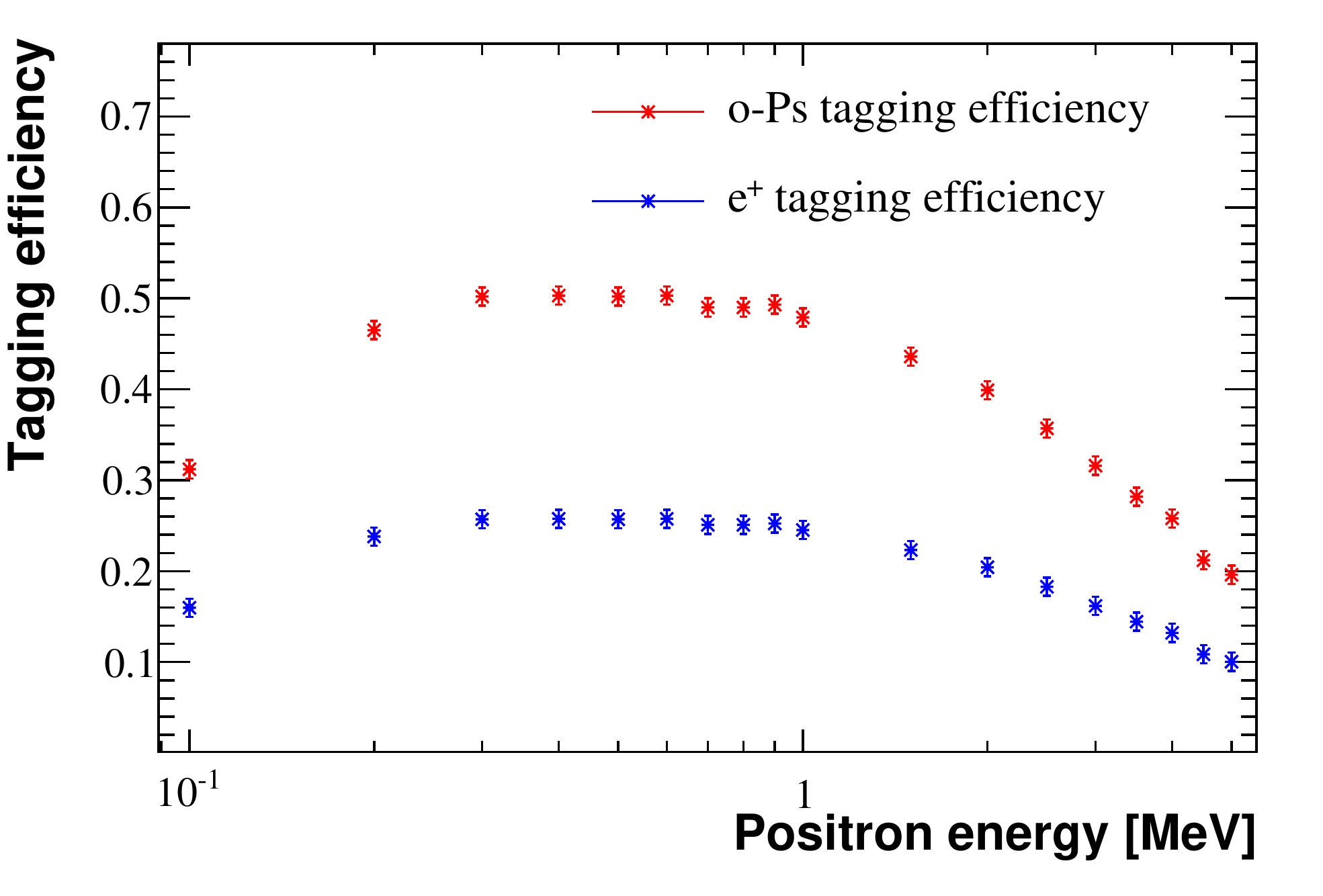}
\caption{Color online. o-Ps tagging efficiency (red) based on the $\rho < \rho_0(E)$ cut, requiring 1\% of direct annihilation contamination fraction, for an ideal 4 m radius detector filled with PC+1.5 g/l PPO. The e$^+$ tagging efficiency (blue) is obtained multiplying the o-Ps tagging efficiency by the o-Ps formation probability.}
\label{fig:tag}
\end{center}
\end{figure}

The pulse shapes differ particularly in the peak positions and widths. We define a discrimination variable, $\rho$, as the ratio between the integrals of the distributions between [0--18] ns and [18--60] ns. An example of $\rho$ distribution for 0.5 MeV positrons is shown in figure \ref{fig:rho}. Even with a not sophisticated estimator, the separation between the two samples is clear: $\rho$ from direct annihilation positrons  is gaussian distributed,  centered in 0.76, while positrons following the \ops\ decay have a broad distribution of $\rho$, depending on the \ops\ life time. Because of the already mentioned energy dependence, we optimized the $\rho <  \rho_0(E)$ cut  by varying the $\rho_0(E)$ threshold with the requirement of  1\%  direct annihilation contamination fraction. Results are shown in figure   \ref{fig:tag}.  The lower efficiency detection at 100 keV, with respect to the one at 500 keV, is due to the detector resolution. 

With this technique, we demonstrate that an event by event tagging of positrons is possible, with   efficiency as high as $\sim$25\% at 0.5 MeV,  equal to the detection efficiency multiplied by the \ops\ formation probability.   More sophisticated algorithms can improve the positronium detection efficiency,  providing a new signature for anti-neutrino experiments and for background rejections in Solar neutrino experiments

\section{Conclusions}

In this paper we measured the \ops\ formation probabilities and life times for the most popular organic liquid scintillators used in neutrino experiments. All the scintillators are characterized by similar \ops\ properties, with $\sim$3 ns mean life and $\sim$50\% formation probability. The distortion of photon emission time distribution induced by the \ops\  intermediate state is a source of systematic in neutrino experiments, since it affects positron event reconstruction and pulse shape discrimination algorithms. 

We further demonstrated  that the \ops\ induced distortion represents a powerful signature for discriminating a fraction of positrons from electrons.  This technique can be exploited  for  both enhancing the anti-neutrino detection efficiency in $\theta_{13}$ and geo-neutrino experiments, and for rejecting   the positron background in Solar neutrino experiments.

\vspace{10 mm}

\begin{acknowledgements}
We are grateful to Ute Schwan, Stefan Schoenert, and Paolo Lombardi  for having provided us the scintillator samples. We thank also Simone Stracka, Pietro Biassoni, and Jos\'e Maneira for useful discussions and comments.
\end{acknowledgements}

%%%%%%%%%%%%%%%%%%%%%%%%%%%%%%%%%%%%%%%
%\bibliography{thebibliography}

\end{document}